\documentclass[sigconf,pbalance]{acmart}

\renewcommand\footnotetextcopyrightpermission[1]{} 
\usepackage[english]{babel}
\usepackage{float}
\usepackage{algpseudocode}
\usepackage{algorithm}

\usepackage{amsmath}
\usepackage{bm}

\usepackage{graphicx}
\usepackage{hyperref}
\usepackage{tikz}
\usepackage{caption}  
\usepackage{subcaption} 
\usetikzlibrary{quantikz}
\usetikzlibrary{decorations.pathreplacing}

\usepackage{enumitem}

\title{Trainability Barriers in Low-Depth QAOA Landscapes}

\author{Joel Rajakumar}
\affiliation{%
  \institution{University of Maryland}
  \city{College Park}
  \state{MD}
  \country{USA}
}
\author{John Golden}
\affiliation{%
  \institution{Information Sciences, Los Alamos National Laboratory}
  \city{Los Alamos}
  \state{NM}
  \country{USA}
}
\author{Andreas B\"artschi}
\affiliation{%
  \institution{Information Sciences, Los Alamos National Laboratory}
  \city{Los Alamos}
  \state{NM}
  \country{USA}
}
\author{Stephan Eidenbenz}
\affiliation{%
  \institution{Information Sciences, Los Alamos National Laboratory}
  \city{Los Alamos}
  \state{NM}
  \country{USA\vspace*{6ex}}
}



\begin{document}

\pagestyle{plain}
\begin{abstract}
The Quantum Alternating Operator Ansatz (QAOA) is a prominent variational quantum algorithm for solving combinatorial optimization problems. Its effectiveness depends on identifying input parameters that yield high-quality solutions. However, understanding the complexity of training QAOA remains an under-explored area. Previous results have given analytical performance guarantees for a small, fixed number of parameters. At the opposite end of the spectrum, barren plateaus are likely to emerge at $\Omega(n)$ parameters for $n$ qubits. In this work, we study the difficulty of training in the intermediate regime, which is the focus of most current numerical studies and near-term hardware implementations. Through extensive numerical analysis of the quality and quantity of local minima, we argue that QAOA landscapes can exhibit a superpolynomial growth in the number of low-quality local minima even when the number of parameters scales logarithmically with $n$. This means that the common technique of gradient descent from randomly initialized parameters is doomed to fail beyond small $n$, and emphasizes the need for good initial guesses of the optimal parameters.
\end{abstract}

\maketitle

\section{Introduction and Background}

The Quantum Alternating Operator Ansatz (QAOA)~\cite{Farhi2014, hadfield_qaoa}, is a leading quantum algorithm designed to solve combinatorial optimization problems.
At its core, QAOA depends on two different quantum operators.  
The first, often referred to as the phase operator, encodes the classical optimization problem being studied. 
The second, known as the mixing operator, facilitates exploration of the solution space by enabling transitions between different quantum states, which in this case represent different solutions to the optimization problem.
By repeatedly applying these operators in an alternating sequence, the algorithm can generate constructive interference between states representing good solutions and destructive interference between states representing poor solutions.
Ideally, QAOA will quickly converge towards a superposition of states representing optimal or near-optimal solutions.

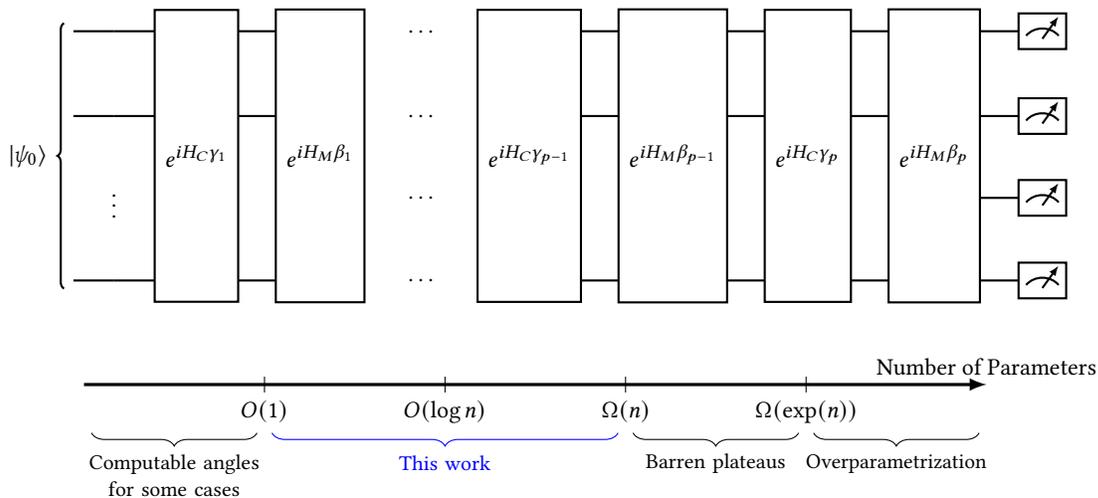
\begin{figure*}[t!]
    \centering
    \begin{quantikz}
        \lstick[4]{$\ket{\psi_0}$}    &\qw    &\gate[4,nwires=3]{e^{iH_C\gamma_1}}        &\gate[4,nwires=3]{e^{iH_M\beta_1}}              &\ \ldots\   &\gate[4,nwires={1,2,3,4}]{e^{iH_C\gamma_{p-1}}}   &\gate[4,nwires=3]{e^{iH_M\beta_{p-1}}} &\gate[4,nwires=3]{e^{iH_C\gamma_p}}  &\gate[4,nwires=3]{e^{iH_M\beta_p}}  &\meter{} \\
        &\qw     &               &         &\ \ldots\     &       &     &       &       &\meter{} \\
        &\vdots           &      &         &\ \ldots\     &       &     &       &       &\meter{} \\
        &\qw            &         &         &\ \ldots\     &       &        &       &       &\meter{} \\
    \end{quantikz}
    \newline
    \begin{tikzpicture}
    \draw[very thick,-{latex}] (0,0) -- (12,0) node[above]{Number of Parameters};
        \foreach \x/\l in {2.4/O(1),4.8/O(\log n),7.2/\Omega(n),9.6/\Omega(\exp(n))}
            \draw (\x,3pt) -- (\x,-3pt) node[below]{$\l$};
        \draw [decorate,decoration={brace,amplitude=5pt,mirror,raise=4ex}]
          (0.1,-0.1) -- (2.3,-0.1) node[align=center,text width=3cm, midway,yshift=-3.5em, font=\small]{Computable angles for some cases};
        \draw [blue, decorate,decoration={brace,amplitude=5pt,mirror,raise=4ex}]
          (2.5,-0.1) -- (7.1,-0.1) node[align=center,text width=2cm, midway,yshift=-3em, font=\small]{This work};
        \draw [decorate,decoration={brace,amplitude=5pt,mirror,raise=4ex}]
          (7.3,-0.1) -- (9.5,-0.1) node[align=center,text width=2cm, midway,yshift=-3em, font=\small]{Barren plateaus};
        \draw [decorate,decoration={brace,amplitude=5pt,mirror,raise=4ex}]
          (9.7,-0.1) -- (11.9,-0.1) node[align=center,text width=2cm, midway,yshift=-3em, font=\small]{\clap{Overparametrization}};
    \end{tikzpicture}
    \caption{Conceptual overview of where this work fits into previous results on the difficulty of training QAOA circuits. For a small, fixed number of parameters, optimal parameters can be computed in some cases~\cite{farhi2015quantum}. When the number of parameters scales linearly in the number of qubits $n$, barren plateaus can emerge~\cite{Larocca_2022}, rendering gradient descent from random parameters exponentially ineffective. Finally, at an exponential number of parameters, overparameterization~\cite{Larocca_2023} makes trainability straightforward, as gradient descent is guaranteed to converge to a global minimum. In this work, we argue that QAOA can suffer from difficult landscapes even in the sublinear regime.}
\label{fig:overview}
\end{figure*}

The exact implementation of the phase and mixing operators depends on a set of parameters that must be tuned for good performance.
In particular, the phase and mixing operators, $U_P$ and $U_M$, are usually implemented as 
\begin{equation}
    U_P = e^{-i\gamma H_C}, U_M = e^{-i\beta H_M}.
\end{equation}
Here $H_C|x\rangle = C(x)|x\rangle$, where $x$ is a binary string representing a solution to the classical optimization problem studied with $C(x)$ being the value of the objective function of solution $x$.
$H_M$, or the mixer Hamiltonian, has several common forms, notably the transverse field mixer $\sum_i X_i$~\cite{Farhi2014} or XY-mixer $\sum_{i,j} X_i X_j + Y_iY_i$~\cite{Wang_2020}.
Taken together with an initial state $\psi_0$ and a set of $2p$ parameters $\bm{\beta} = \{\beta_1, \ldots, \beta_p\}$ and $\bm{\gamma} = \{\gamma_1, \ldots, \gamma_p\}$, these operators form a $p$-round QAOA in the following way:
\begin{equation}
    |\psi_p(\bm{\beta}, \bm{\gamma})\rangle = e^{-i \beta_p H_M} e^{-i \gamma_p H_C} \dots e^{-i \beta_1 
    H_M} e^{-i \gamma_1 H_C} |\psi_0\rangle.
\end{equation}
The goal is to find $\bm{\beta}, \bm{\gamma}$ such that sampling from $|\psi(\bm{\beta}, \bm{\gamma})\rangle$ will return states representing good solutions (as measured by the values of $C(x))$ with high probability. In this work, we treat the hyperparameter $p$ as an architecture-agnostic way of capturing the scaling of circuit depth.

Determining parameters that lead to good QAOA performance is a significant outstanding research area.
For some problem classes, analytical descriptions of good parameters (along with performance guarantees) exist, e.g. for $p=1$~\cite{farhi2015quantum} or $n\to\infty$~\cite{boulebnane2024solving}.
In the general case, QAOA is usually implemented as a hybrid classical-quantum heuristic.
In this framework, a quantum computer (or simulator) is used to estimate 
\begin{equation}
    \langle H_C\rangle_{\bm{\beta},\bm{\gamma}} \equiv \langle \psi_p(\bm{\beta}, \bm{\gamma})|H_C|\psi_p(\bm{\beta}, \bm{\gamma})\rangle
\end{equation}
for an initial set of parameters $\{\bm{\beta}, \bm{\gamma}\}$.
A classical optimization algorithm, e.g. gradient descent, is then used to update $\{\bm{\beta}, \bm{\gamma}\}$, which are then passed to the quantum device and evaluated.
This process is then repeated, potentially as a subroutine in a more complex classical optimization algorithm, e.g. basin hopping~\cite{Wales_1997}.

While this heuristic approach has been used extensively to study the potential efficacy of QAOA~\cite{Lotshaw_2021, Golden_2023}, little is concretely known about the computational hardness of the classical optimization routine (beyond the general NP-hardness of training variational quantum algorithms~\cite{Bittel_2021}), particularly in the low $p$ regime.
In the regime of $p=\Omega(n)$, QAOA is likely to suffer from barren plateaus~\cite{Larocca_2022}.
With barren plateaus, the average variance of the gradient of the cost function is suppressed as $n$ increases. 
In practice, this means that gradient descent from a random set of parameters is exponentially unlikely to lead to high-quality solutions.
However, when $p$ is exponentially larger than $n$ an effect known as overparameterization makes discovering good parameters straightforward~\cite{Larocca_2023}.
In this case, the large degree of expressibility in the ansatz allows gradient descent to escape the barren plateaus and navigate to a globally optimal solution with near certainty.
While this is not an exhaustive list of theoretical results for the difficulty of learning (see e.g.~\cite{Anschuetz_2022}, which only applies for specific circuit models, or~\cite{boulebnane2024solving} which studies QAOA for strict satisfiability), it captures the general state of affairs. See Fig.~\ref{fig:overview} for a conceptual overview.

\section{Outline of Results}
In this work, we characterize the difficulty of parameter-finding in the regime $1<p<O(n)$, filling in the gap between the theoretical results at $p=1$ and $p=\Omega(n)$.
This regime is particularly important as it is where almost all current numerical simulations and near-term implementations of QAOA are done~\cite{Preskill_2018,Golden_2023, Lykov_2023, Golden_2023_Numerical}.
In particular, we focus on the common parameter finding technique of gradient descent from random initialization~\cite{Lotshaw_2021}, and we numerically study how the cost landscape of QAOA circuits with a sublinear number of parameters changes with $n$ and $p$. 
We study the standard problem of MaxCut on Erd\H{o}s-R\'enyi random graphs with edge probability $0.5$, which is known to exhibit barren plateaus~\cite{Larocca_2022} and overparameterization~\cite{Larocca_2023}.

To assess the difficulty of navigating the cost landscape for QAOA in this regime, we study the quality and quantity of local minima. 
The quality of local minima -- which we define precisely in Section~\ref{sec:quality} -- captures the fraction of the cost landscape in which gradient descent reaches high-quality solutions.
Meanwhile, the quantity of local minima -- which we estimate using techniques described in Section~\ref{sec:quantity} -- characterizes the effectiveness of local search methods to escape suboptimal local minima, e.g. random restarts, momentum, simulated annealing, or basin hopping. The quantity of minima also captures global properties of the landscape such as ruggedness and degeneracy of minima.

We study how the quality and quantity of local minima scale under three conditions: fixed $n$ and increasing $p$, fixed $p$ and increasing $n$, and $p \propto \log n$.
In each case, we develop numerical evidence that local optimization methods will fail to find high-quality minima without a good initial guess of optimal parameters.
In particular, we find
\begin{enumerate}
    \item Exponential decay in the fraction of the landscape from which gradient descent reaches high-quality solutions at $p = O(\log(n))$.
    \item Super-polynomial growth in the number of local minima at $ p = \Omega(\log(n))$.
\end{enumerate}
Note that throughout this paper we use $O(f(n))$ to refer to functions of $n$ which are asymptotically less than or equal to $f(n)$ and $\Omega(f(n))$ for functions which are asymptotically greater than or equal to $f(n)$. We note that a majority of our numerics are in the regime where $p$ is comparable to $n$ and $n \leq 16$, so we can only develop evidence towards asymptotic scaling, and not directly prove it.

These results strongly suggest that in the case of MaxCut on Erd\H{o}s-R\'enyi graphs, QAOA parameter finding with local optimization from a random initialization is at least superpolynomially difficult. This emphasizes the need for more intelligent means of parameter finding for QAOA. 
Examples of such approaches include warm-starts, i.e. a smart guess for either the initial state $|\psi_0\rangle$ or initial set of parameters $\bm{\beta}, \bm{\gamma}$, or developing and exploiting a more robust understanding of the global cost landscape, e.g.~\cite{choy2024energy}.

\begin{figure*}[ht!]
    \begin{subfigure}[b]{0.29\textwidth}
        \includegraphics[width=0.9\textwidth]{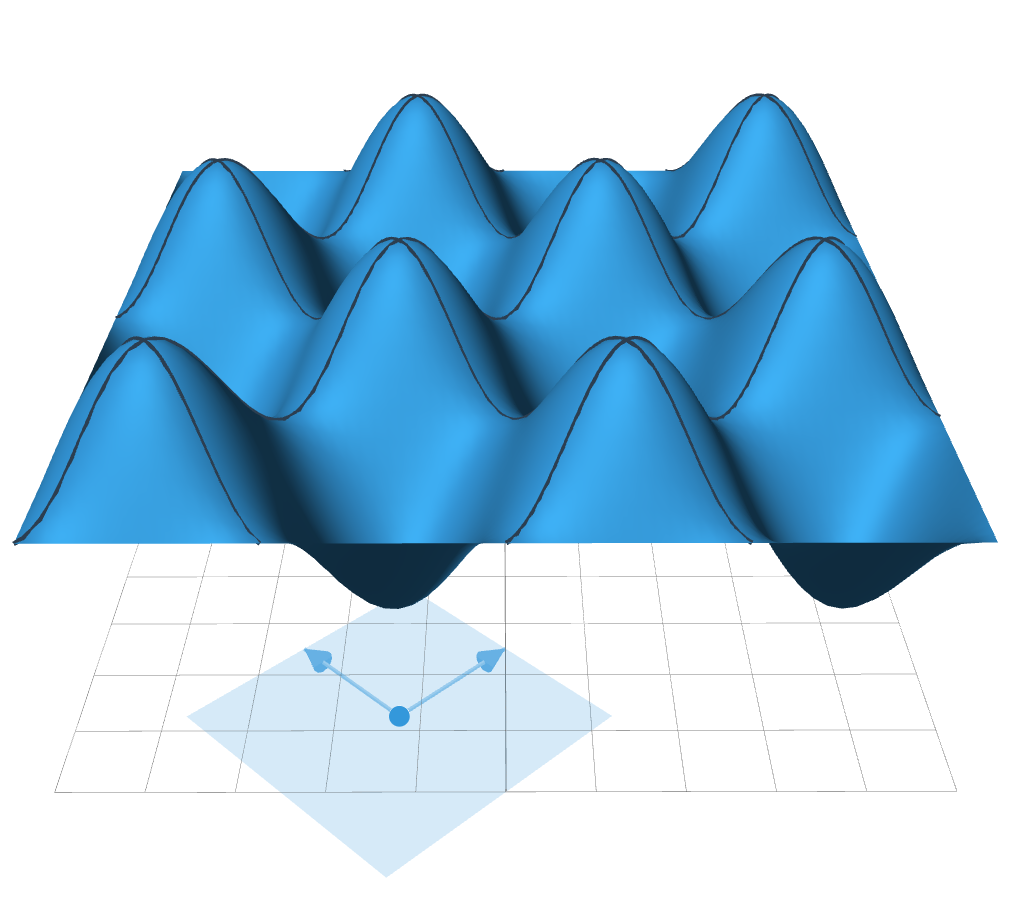}
        \caption{Division of cost landscape into\newline basins corresponding to local minima}
        \label{fig:Minima}
    \end{subfigure}
    \hfill
    \begin{subfigure}[b]{0.35\textwidth}
        \includegraphics[width=\textwidth]{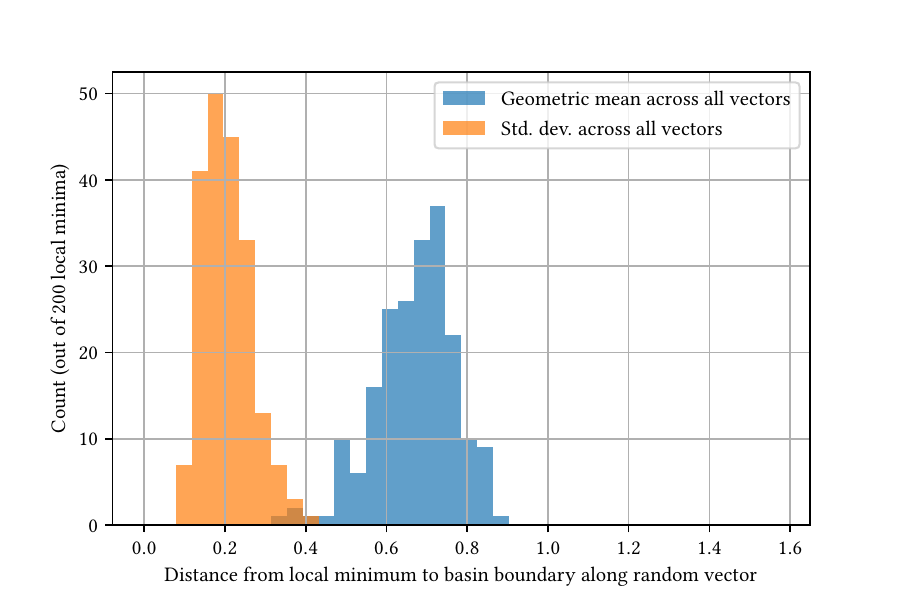}
        \caption{Variation in distance to basin boundary\newline across choice of random vector}
        \label{fig:Vector_STD}
    \end{subfigure}
    \hfill
    \begin{subfigure}[b]{0.35\textwidth}
        \includegraphics[width=\textwidth]{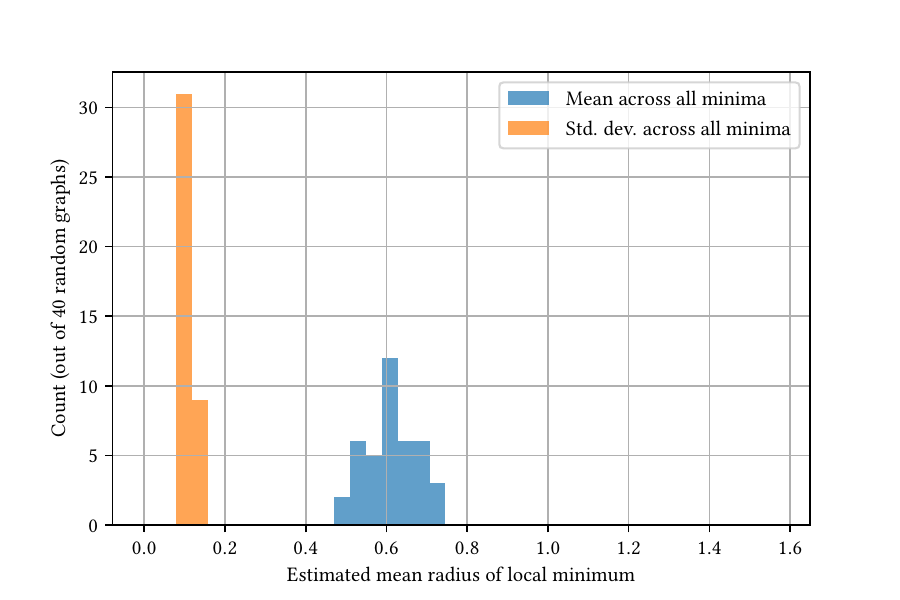}
        \caption{Variation in estimated radius\newline across choice of local minima}
        \label{fig:Minima_STD}
    \end{subfigure}
    \caption{\textbf{Estimating the quantity of local minima using mean radius of basins.} In figure (a), we depict a cost landscape with many local minima. We show a division of the cost landscape into basins (black lines) and highlight the boundaries of a particular basin (light blue). The number of minima is estimated by dividing the total volume of parameter space by the average volume of the basins. In figure (b) and (c), we sample Erd\H{o}s-R\'enyi random graphs with edge probability $0.5$ and $n=8$. We then perform gradient descent from 200 random initializations of parameters (with number of rounds $p=5$) to arrive at 200 local minima. The radius of each minimum's basin is estimated by using Algorithm~\ref{alg:estimation}. We observe that the radius estimates show low variation across choice of random vector and choice of minima (which are randomly selected in Algorithm~\ref{alg:estimation}). This supports the fact that the mean basin radius is an effective proxy for the overall quantity of local minima.}
    \label{fig:basins}
\end{figure*}

\section{Methods}\label{sec:methods}

For our numerical experiments, we study the common QAOA test case of solving MaxCut on Erd\H{o}s-R\'enyi random graphs with edge probability $0.5$.
The objective of MaxCut is to find a partition of the graph that maximizes the number of edges crossing the partition,
and for a graph with edge set $E$ it is described by the cost Hamiltonian 
\begin{equation}
    H_C = \sum_{(i,j) \in E} \frac{1-Z_i Z_j}{2}.
\end{equation}
The standard mixer for MaxCut is the transverse-field mixer~\cite{Farhi2014}
\begin{equation}
    H_M = \sum_{i=1}^n X_i.
\end{equation}
As these Hamiltonians have integer-valued coefficients, they have fixed periods of at most $2\pi$, i.e. $e^{iH_M (\beta + 2\pi)} = e^{iH_M\beta}$. The combination of the MaxCut cost Hamiltonian and the transverse-field mixer also has additional symmetries in $\beta$ and $\gamma$ which further reduces the parameter space to be explored~\cite{Lotshaw_2021}. However, our numerical methods capture the density of local minima in the cost landscape, so the exact bounds of the parameter space do not matter for our results.

There are many existing metrics to evaluate the output of a QAOA circuit. In the context of the QAOA cost landscape of combinatorial optimization problems, we choose to study the approximation ratio, which is the ratio of the expected quality of the solution found by QAOA and the optimal solution. It is defined as
\begin{align}
    \text{Approx. Ratio}(\beta,\gamma) = \frac{\langle H_C\rangle_{\bm{\beta},\bm{\gamma}}}{\max_{x \in \{0,1\}^n} \bra{x} H_C \ket{x}} 
\end{align}

The approximation ratio is well-suited for maximization problems such as MaxCut, which have $C(x) \ge 0~\forall x \in \{0,1\}^n$. For other problem types, e.g. minimizing the energy of an Ising model, other metrics are more commonly used, such as the absolute difference between the expectation value and the true minimum, or the probability of sampling the true minimum when measured in the computational basis \cite{Willsch_2020}. However, the absolute value of the true minimum and the number of possible computational basis states vary along with $n$, which makes these alternative metrics difficult to compare when $n$ is scaled.

The approximation ratio is computed by performing an exact, noise-free state-vector simulation of the QAOA circuit with parameters $\{\bm{\beta},\bm{\gamma}\}$, and directly computing the expectation value of the problem cost function at the output state. We use the \texttt{JuliQAOA} package~\cite{Golden_2023_Numerical} for our numerical simulations. 

\begin{figure*}[ht!]
    \begin{subfigure}[b]{0.48\textwidth}
        \includegraphics[width=\textwidth]{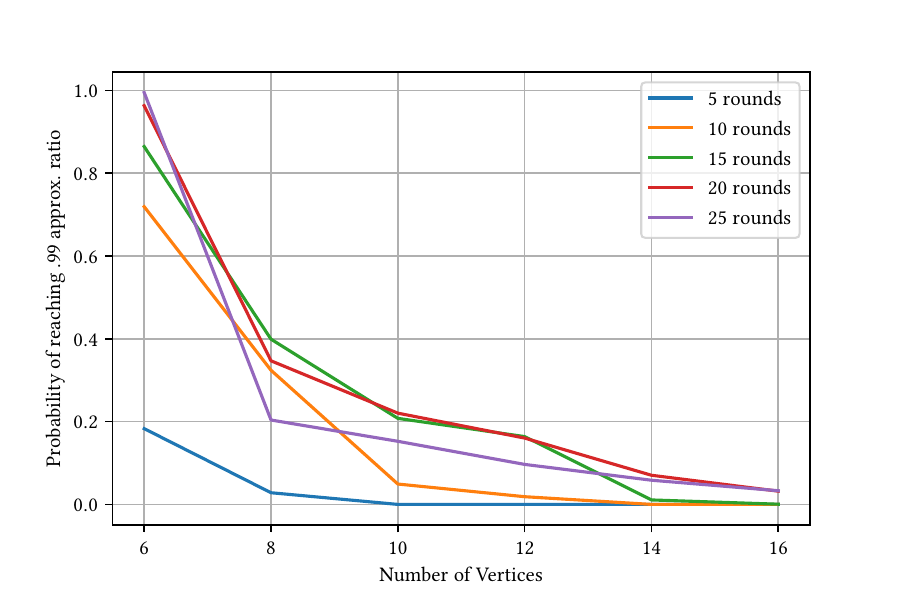}
        \caption{Quality of Local Minima}
        \label{fig:quality_fixedp}
    \end{subfigure}
    \hfill 
    \begin{subfigure}[b]{0.48\textwidth}
        \includegraphics[width=\textwidth]{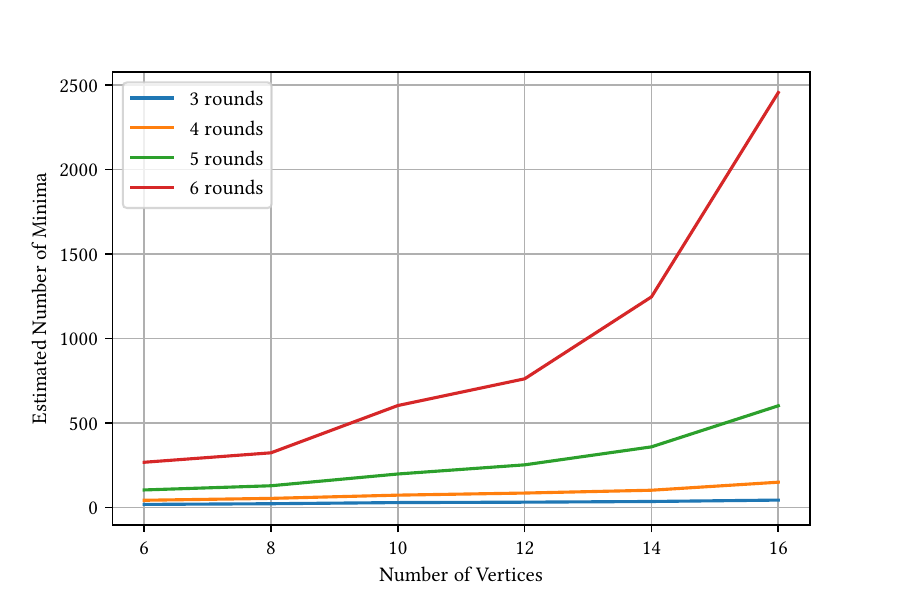}
        \caption{Quantity of Local Minima}
        \label{fig:quantity_fixedp}
    \end{subfigure}
    \caption{\textbf{Scaling in $n$ of local minima quality and quantity for fixed $p$}. For each $n$, we sample 40 Erd\H{o}s-R\'enyi random graphs with edge probability $0.5$, and for each graph, we sample 200 random initialization points in the cost landscape. Gradient descent is performed from each of these points to result in 200 (possibly non-distinct) local minima. (a) The quality of local minima is determined by the fraction of these local minima which reach an approximation ratio $ > 0.99$, and (b) the quantity of minima is estimated by applying Algorithm~\ref{alg:estimation} for each local minimum and taking the average.}
    \label{fig:fixedp}
\end{figure*}

\subsection{Evaluating the Quality of Local Minima}\label{sec:quality}
To study the quality of local minima in the QAOA cost landscape of a particular graph problem instance, we uniformly sample points from the parameter space and perform gradient descent (via the Broyden-Fletcher-Goldfarb-Shannon algorithm~\cite{bfgs}) from each one. For each graph, we compute the probability that this technique results in an approximation ratio above a fixed cutoff of 0.99. We choose a high cutoff of 0.99 to expose large variations in the cost landscape with $n$ and $p$. However, the particular choice of cutoff should not matter asymptotically when $n$ and $p$ are scaled (as long as the cutoff is sufficiently far above the mean value of the approximation ratio over all $\bm{\beta},\bm{\gamma}$).

\subsection{Estimating the Quantity of Local Minima}\label{sec:quantity}

Many existing methods to estimate the number of minima in a landscape are practically infeasible in the QAOA setting. For instance, performing a fine-grained grid search becomes difficult past $p>2$ since the search space increases exponentially. Other sampling methods from literature \cite{10.1162/EVCO_a_00100} require a number of samples which grows with the number of local minima, which again becomes difficult for large $p$. 

Instead, we study the size of the `basin of attraction' of local minima as a proxy for the number of minima. The basin of attraction of a local minimum is the region of the cost landscape in which deterministic gradient descent (such as BFGS) will lead to the same local minimum. The entire cost landscape can be divided into different basins of attractions for different local minima. We depict such basins for an example case in Figure~\ref{fig:basins}. Basin sizes give a sense of how far apart local minima are from each other, and when combined with the overall volume of the parameter space, we can estimate the total number of local minima.

In order to estimate the mean basin size, and thus total number of minima, we employ a novel algorithm, described here and in detail in Algorithm~\ref{alg:estimation}. If $V$ is the product of the ranges of the parameters (i.e. the total volume of the parameter space), and $v$ is the average volume of the basins of attraction, $V/v$ is an estimate of the number of local minima. If each basin is modeled as a $p$-ball in the landscape, an estimate of its volume can be obtained by choosing $p$ orthogonal vectors originating from the local minimum and estimating the distance from the minimum to the boundary of the basin along each vector.  An estimate of this distance is obtained using a binary search to find the closest point along the vector such that gradient descent will land in a different local minimum. The estimated radius of the basin loosely captures how far apart local minima are (in Section~\ref{sec:log}, we directly study the estimated radius and the implications of its variations).

\begin{algorithm}[b]
\caption{Number of Minima Estimation}\label{alg:estimation}

\begin{algorithmic}
\For{$i \in samples$}
\State $x\gets $ BFGS(random.vector($num\_params$))
\State $vectors \gets $ random.orthogonal($num\_params,num\_params$)
\For{$vector \in vectors$}
\State $l \gets 0, r \gets 2\pi$
\While{$r-l > precision$}
\State $radius \gets \frac{l+r}{2}$
\State $x' \gets $ BFGS($x + vector*radius$)
\If{$\|x'-x\| < \epsilon $}
\State $l \gets radius$
\Else
\State $r \gets radius$
\EndIf
\EndWhile
\State $radii \gets radius$
\EndFor
\State $mean\_radius \gets $geometric\_mean ($radii$)
\State $volumes \gets $ volume.p\_ball($mean\_radius$)
\EndFor
\State $num\_minima \gets V/$mean($volumes$)\\
\Return $num\_minima$
\end{algorithmic}
\end{algorithm}

\begin{figure*}[ht!]
    \begin{subfigure}[b]{0.48\textwidth}
        \includegraphics[width=\textwidth]{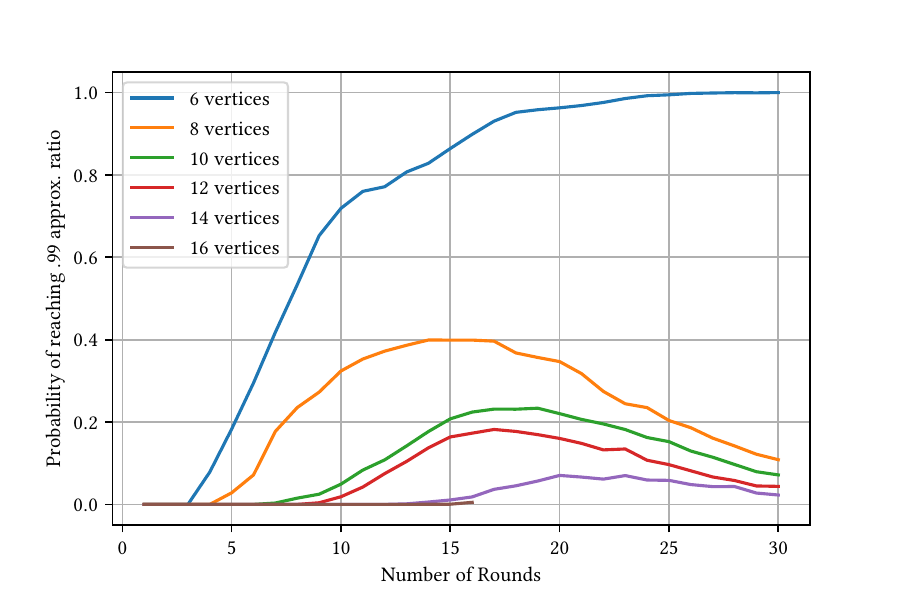}
        \caption{Quality of Local Minima}
        \label{fig:quality_fixedn}
    \end{subfigure}
    \hfill
    \begin{subfigure}[b]{0.48\textwidth}
        \includegraphics[width=\textwidth]{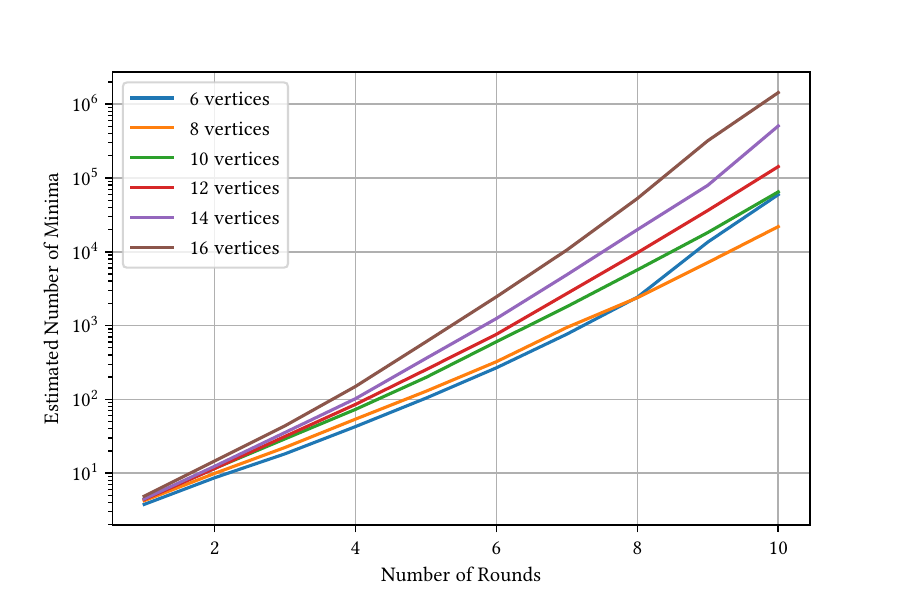}
        \caption{Quantity of Local Minima}
        \label{fig:quantity_fixedn}
    \end{subfigure}
    \caption{\textbf{Scaling in $p$ of local minima quality and quantity for fixed $n$}. For each $n$, we sample 40 Erd\H{o}s-R\'enyi random graphs with edge probability $0.5$, and for each $p$, we sample 200 random initialization points for gradient descent on each graph. Local minima quality is determined by the fraction of uniformly sampled points from which gradient descent reaches an approximation ratio $ > 0.99$, and number of minima is estimated using Algorithm~\ref{alg:estimation}. }
    \label{fig:fixedn}
\end{figure*}

We numerically find that the estimated radius of a local minimum's basin in Algorithm~\ref{alg:estimation} shows relatively low variance across choice of minima and the estimated distance from a minimum to its basin's boundary shows low variance across choice of random vector (see Figure~\ref{fig:basins}). This is significant because it validates both of our experimental methods as follows: 

\textbf{Quality of Local Minima} -- the low variance across choice of minima provides evidence that basins of local minima are largely homogeneously distributed throughout the cost landscape. If this were not so, some of the estimated radii would be significantly lower where there are clusters of local minima. Crucially, this means that our uniform sampling of initialization points for gradient descent in Section~\ref{sec:quantity} reflects a mostly uniform sampling of the local minima in the landscape. Therefore, when we study the fraction of random initializations that reach a cutoff approximation ratio under gradient descent, we get a direct estimate of the fraction of local minima with good approximation ratios.

\textbf{Quantity of Local Minima} -- the low variance across choice of vector implies that the boundary of a minimum's basin is roughly the same distance from the minimum regardless of which direction from the minimum is explored. This provides evidence that each local minimum is located roughly at the center of its basin and that each basin is accurately modeled as a symmetric $p$-ball around the local minimum. The low variance across minima indicates that each basin is of similar size, so the estimate of the average volume of local minima basins is close to the real value.

\section{Cost Landscape with Scaling System Size} \label{sec:fixedp}

We begin by studying how the quality and quantity of local minima scale with $n$ for fixed $p$. We observe that gradient descent starts to perform significantly worse as $n$ increases (see Figure~\ref{fig:quality_fixedp}). In Section~\ref{sec:log}, we provide evidence that this observed decay in success probability is likely to be exponential in $n$ for fixed $p$. Our results on the number of minima (see Figure~\ref{fig:quantity_fixedp}) suggest that the landscape is becoming increasingly rugged with $n$ as more minima crop up in the landscape. This provides evidence that when there are not enough parameters, gradient descent often gets stuck in suboptimal minima \cite{Anschuetz_2022, Larocca_2023}. Thus even for low, fixed $p$, parameter finding is potentially exponentially difficult when using random initialization.

One of the most promising avenues away from this issue is a heuristic commonly referred to as parameter transfer~\cite{Brando2018ForFC,Wurtz,Galda,Shaydulin_2023}. This warm-start technique initializes gradient descent with the average of optimized parameters found for graph problems on smaller $n$. This approach potentially allows one to bootstrap optimized parameters to larger $n$ while keeping $p$ fixed, and has shown some modest success up to 127 qubits~\cite{Pelofske2023ScalingWQ}. While there are some strong theoretical arguments in favor of parameter transfer~\cite{Brando2018ForFC}, it has yet to be proven as a means of consistently scaling QAOA performance up in $n$.

\section{Cost Landscape with Scaling\linebreak Number of Parameters}
Next, we study how the quality and quantity of local minima scale with $p$ for fixed $n$. The addition of parameters to the cost landscape can have two effects on the number of minima. The new dimensions can turn existing local minima into saddle points \cite{PhysRevA.107.062404}, thereby reducing the number of local minima. Or, new minima can crop up in higher dimensions because the size of the parameter space is exploding exponentially with $p$ \cite{Fontana_2022, Larocca_2023}. Our results (see Figure~\ref{fig:quantity_fixedn}) indicate that the second effect wins out, at least in the shallow $p$ regime we numerically explore. 

However, an interesting phenomenon we observe is that many of these minima start to show good approximation ratios at an intermediate range of $p$ (see the peaks in Figure~\ref{fig:quality_fixedn}). We hypothesize that this is due to the first effect, where increasing dimensions allows many suboptimal minima to turn into saddle points that lead to better local minima (albeit a very large number of them). We note however that the peak decays as $n$ is increased and the start of the peak requires larger and larger $p$ as $n$ is increased ($p > n$). This provides evidence that training could remain difficult at large $n$, even if $p$ is allowed to increase sublinearly with $n$. We further investigate this in Section~\ref{sec:log} when $p$ is allowed to scale logarithmically with $n$.

\begin{figure*}[ht!]
    \begin{subfigure}[b]{0.48\textwidth}
        \includegraphics[width=\textwidth]{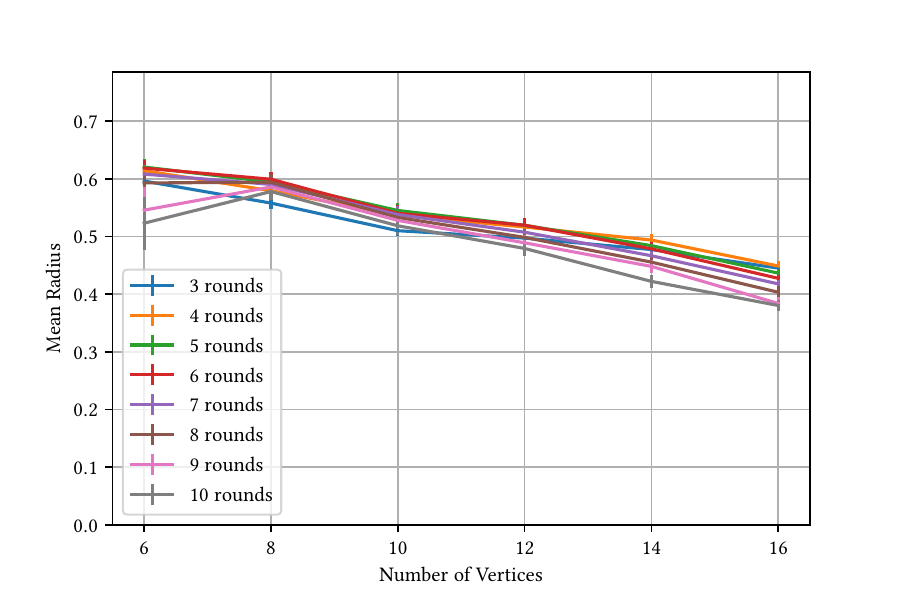}
        \caption{Mean Radius of Local Minima scaling with $n$}
        \label{fig:radius_fixedp}
    \end{subfigure}
    \hfill
    \begin{subfigure}[b]{0.48\textwidth}
        \includegraphics[width=\textwidth]{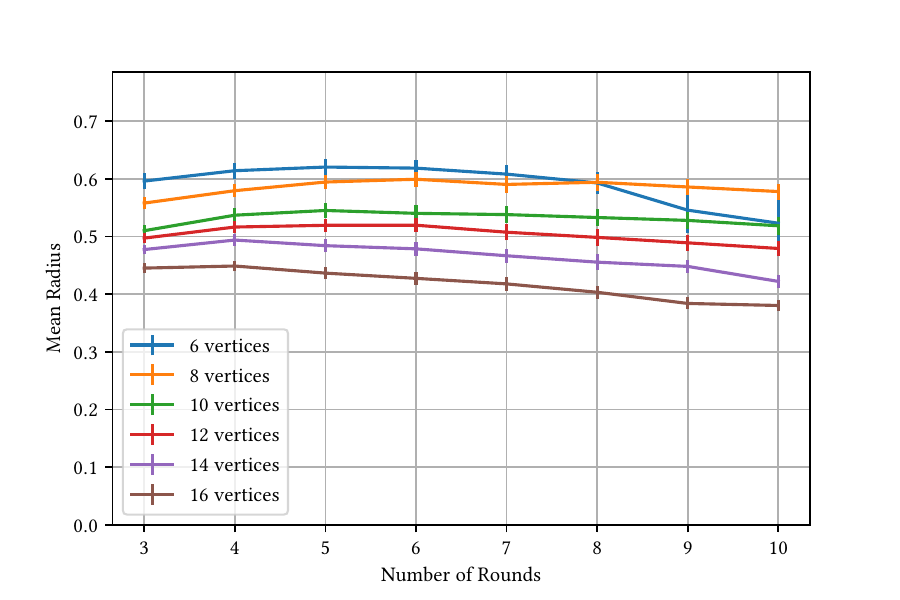}
        \caption{Mean Radius of Local Minima scaling with $p$}
        \label{fig:radius_fixedn}
    \end{subfigure}
    
    \caption{\textbf{Scaling in $n$ and $p$ of mean radius of minima.} For each $n$, we sample 40 Erd\H{o}s-R\'enyi random graphs with edge probability $0.5$, and for each $p$, we sample 200 random initialization points for gradient descent on each graph. The fact that the size (a) decreases as $n$ increases and (b) stays roughly constant in $p$ suggests that the total number of local minima scales superpolynomially for $p\propto \log(n)$.}
    \label{fig:radius}
\end{figure*}

For each $n$, when $p$ is increased past some fixed point, the minima start to perform poorly again. This suggests that the growth in number of poor minima begins to outpace the growth in number of good minima, which is likely evidence of the onset of barren plateaus \cite{McClean_2018}. In practice, the heuristic of greedy recursive initialization \cite{Golden_2023, PhysRevA.107.062404} can be effective in this regime of large $p$. This technique involves initialization of gradient descent on larger $p$ with optimized parameters found for smaller $p$, effectively bootstrapping in $p$.
However, as with parameter transfer, the exact efficacy of greedy recursive initialization is still an open question.

\section{Cost Landscape with Scaling Parameters\linebreak and System Size} \label{sec:log}

Our results demonstrate that the cost landscape exhibits several interesting properties when $p$ and $n$ are scaled. A natural question is to determine whether there is a `sweet spot' scaling of $p$ with $n$ such that $p$ is small enough to avoid barren plateaus, and yet large enough to avoid suboptimal local minima with enough expressibility. We investigate $p \propto \log(n)$ in particular because in \cite{Golden_2023}, at least $\Omega(\log(n))$ parameters were shown to be required to reach good approximation ratios for a variety of combinatorial optimization problems, even when using state-of-the-art heuristics such as recursive greedy extrapolation and basin-hopping. Moreover, near-term implementations of quantum circuits are limited to $O(\log n)$ depth due to the build-up of noise.

Our results indicate that when $p \propto \log(n)$ and $n$ is increased, there is a superpolynomial growth in the number of minima, and an exponential decay in the quality of these minima.

\begin{figure}[ht!]
    \centering
    \includegraphics[width=\linewidth]{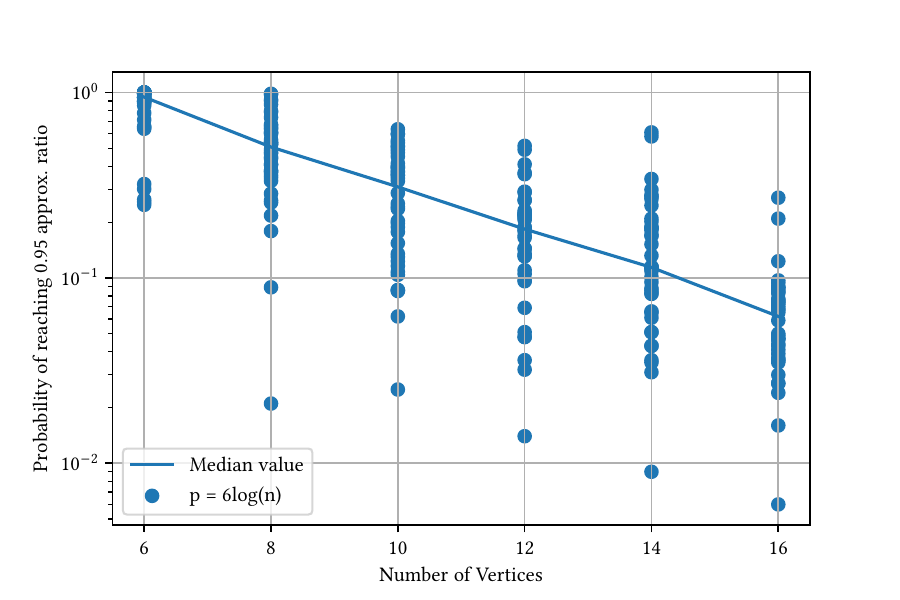}
    \caption{ \textbf{Quality of local minima at $p \propto \log n$. } For each $n$, we sample 40 Erd\H{o}s-R\'enyi random graphs with edge probability $0.5$, set $p = 6 \log n$, and sample 1000 random initialization points for gradient descent on each graph. Each point in the plot represents the fraction of random initializations from which gradient descent reaches an approximation ratio of at least 0.95 for a single graph. Our results indicate that the probability of reaching high-quality minima with gradient descent decays exponentially with $n$. Moreover, the observed exponential decay in the probability of reaching an approx. ratio of $0.95$ implies a general exponential decay in the probability of reaching any cutoff $c > 0.95$.}
    \label{fig:logn}
\end{figure}

\textbf{Quality of Local Minima} -- we choose $p =6\log n$ in our numerics as this corresponds most closely to the peak observed for $n=8$ in Figure~\ref{fig:quality_fixedn}. We find an exponential decay in the probability of finding good minima when $n$ is scaled (see Figure~\ref{fig:logn}). We also note that, if we label the round with the peak quality in~\ref{fig:quality_fixedn} as $p_\text{peak}$, then $p_{\text{peak}}$ seems to grow faster in $n$ than $6\log n$. This is compounded by the fact that the height of the peak itself decays with $n$. For these reasons, we suspect that exponential decay in the quality of minima will be observed when $n$ is scaled regardless of the coefficient in $p = c \log n$.

\textbf{Quantity of Local Minima} -- it is numerically difficult to directly distinguish polynomial from superpolynomial (but non-exponential) growth simply by examining the estimated number of minima. Instead, we directly study the estimated mean radius of the basins of the local minima as a proxy for the number of minima. The intuition for showing superpolynomial growth at $p = \Omega(\log n)$ is as follows. If the basins have the same mean radius as $n$ and $p$ are increased, then a landscape of $p=\log(n)$ parameters would have $O(n)$ such basins (polynomial growth in the number of minima). If the mean radius decreases as $n$ is increased and does not increase as $p$ is increased, then there must be a superpolyomial number of minima when $p = \log n$. Explicitly, suppose the mean radius of the basins decays as $O(f(n))$. This means that their volumes are roughly $O(f(n)^p)$. At $p= \log n$, the number of minima is $\Omega(1/f(n)^{\log n}) = \Omega(n^{\log(1/f(n))})$. This is a superpolynomial lower bound in $n$ for any decreasing $f(n)$. Our numerical experiments support superpolynomial growth in the number of minima, as we find that the mean radius decays with $n$ (See Figure~\ref{fig:radius_fixedp}) and the mean radius does not increase with $p$ (See Figure~\ref{fig:radius_fixedn}). 

These observations have implications that apply beyond the regime of just $p \propto \log(n)$. One implication is that the quality of local minima must decay exponentially for any $p = O(\log n)$, including $p$ fixed to a constant as in Section~\ref{sec:fixedp}. This is because in the regime of $p < n$, the quality of minima strictly decreases when $p$ is decreased (see Figure~\ref{fig:quality_fixedn}). Similarly, there must be a superpolynomial growth in the number of minima for any $p = \Omega(\log n)$ because the quantity of minima strictly increases as $p$ is increased (see Figure~\ref{fig:quantity_fixedn}). Crucially, our results demonstrate that the exponential decay in the quality of local minima and the superpolynomial blowup in the quantity of local minima occur in an overlapping regime of $p \propto \log n$. And since quality decays exponentially for $p$ at or below $\log n$, while quantity increases superpolynomially for $p$ at or above $\log n$, there is no `sweet spot' which avoids these two issues.

\section{Conclusion}
Our results provide numerical evidence that QAOA landscapes can be difficult to optimize over, even when the number of parameters is sublinear in $n$. This means that good initial guesses of optimal parameters are required in order to start gradient descent in a region of the cost landscape that does not encounter too many suboptimal local minima. Several heuristics achieve this by initializing gradient descent with optimal parameters found for lower $n$ and $p$. Other techniques that are used in practice involve restricting the parameter space to small subsections that are analytically guaranteed to contain good solutions \cite{basso2022, Sureshbabu_2024}. It would be interesting to understand whether a superpolynomial number of suboptimal minima emerge in the subspace of the cost landscape that such heuristics target.

Our results also represent some of the first experiments on the quality of solutions obtained with QAOA trained by gradient descent, when $n$ is scaled. Most existing numerical results have involved scaling with $p$ and focus on $p = \Omega(n)$ with $n < 10$. Our experiments at large system size are enabled by the recent development of high-performance QAOA simulators such as \cite{Golden_2023_Numerical, Lykov_2023}. Our observations showcase a gap in the theory of QAOA 
trainability and invite further theoretical analysis of circuits with a sublinear number of parameters. For example, is there an analytical reason behind the experimentally observed decay in the success probability with $n$? 

Our results also showcase the value of studying QAOA performance `experimentally.' To our knowledge, we are the first to observe a peak in the performance of QAOA with increasing $p$ (Figure~\ref{fig:fixedn}). This observation invites several more open questions. Does this peak persist for constrained problems (e.g. $k$-Densest Subgraph) or graphs with lower connectivity (e.g. MaxCut on $k$-regular graphs)? We experimentally observe that the peak decays with increasing $n$ and $p$. We offer barren plateaus as a possible explanation for this phenomenon, but the connection is not concrete. Barren plateaus are generally observed in the asymptotic scaling of gradients with $n$ and $p$. If the decay of the peak is indeed due to barren plateaus, then our results would serve as experimental confirmation of its effects on approximation ratios obtained by gradient descent for specific values of $n$ and $p$. We leave further exploration of the onset of barren plateaus to future work.

Besides $p$, several other hyperparameters affect the performance of QAOA, including choice of mixer, ansatz, and initial state. Further numerical investigation of trainability under these hyperparameters is also needed. 

\section{Acknowledgements}
We thank Marco Cerezo for many helpful discussions on these and related topics. This material is based upon work supported by the National Science Foundation Graduate Research Fellowship Program under Grant No. DGE 1840340. This work was supported by the U.S. Department of Energy (DOE) through a quantum computing program sponsored by the Los Alamos National Laboratory (LANL) Information Science \& Technology Institute. Research presented in this article was supported by the NNSA’s Advanced Simulation and Computing Beyond Moore’s Law Program at Los Alamos National Laboratory. This material is based upon work supported by the U.S. Department of Energy, Office of Science, National Quantum Information Science Research Centers, Quantum Science Center. This research used resources provided by the Darwin testbed at Los Alamos National Laboratory (LANL) which is funded by the Computational Systems and Software Environments subprogram of LANL’s Advanced Simulation and Computing program (NNSA/DOE). 

This work has been assigned LANL technical report number LA-UR-24-20039.

\bibliographystyle{plainurl}
\bibliography{references-arxiv}

\end{document}